\documentclass[letterpaper]{article} % DO NOT CHANGE THIS
\usepackage[preprint]{aaai2027}  % Named version; use [submission] for anonymous review
\usepackage[hyphens]{url}  % DO NOT CHANGE THIS
\usepackage{graphicx} % DO NOT CHANGE THIS
\usepackage{natbib}  % DO NOT CHANGE THIS AND DO NOT ADD ANY OPTIONS TO IT
\usepackage{caption} % DO NOT CHANGE THIS AND DO NOT ADD ANY OPTIONS TO IT

\usepackage{amsmath}
\usepackage{listings}
\usepackage{xcolor}
\usepackage{fvextra}
\usepackage{makecell}
\usepackage{placeins}
\usepackage{flafter}
\usepackage{tabularx}
\usepackage{booktabs}

  \usepackage[most]{tcolorbox}
  \newtcolorbox{finding}{
    colback=black!6, colframe=black!6, boxrule=0pt, sharp corners,
    left=6pt, right=6pt, top=5pt, bottom=5pt,
    before skip=8pt, after skip=8pt}

\newcommand{\etal}[1]{et~al.}

\newcommand{\gopi}[1]{\textcolor{red}{{\it [Gopi: #1]}}}
\newcommand{\rqthree}{RQ3: Can Current Models Reliably Execute Deletion-Only Edits?}

\newcommand{\showpagenumbers}{%
  \pagestyle{plain}%
  \thispagestyle{plain}%
}
\title{To Add Is Machine, To Delete Is Human: \\Measuring and Mitigating Deletion Avoidance in LLM Code Editing}
\author{
Amir M. Ebrahimi,
Mohammed Mehedi Hasan,
Aaditya Bhatia,\\
Gopi Krishnan Rajbahadur,
Ahmed E. Hassan
}
\affiliations{
School of Computing, Queen's University, Kingston, Ontario, Canada\\
amir.ebrahimi@cs.queensu.ca,
mohammedmehedi.hasan@queensu.ca,\\
aaditya.bhatia@cs.queensu.ca,
grajbahadur@acm.org,
ahmed@cs.queensu.ca
}

\begin{document}

\maketitle
\showpagenumbers

\begin{abstract}
Large language models increasingly write and repair production code, yet evidence is mounting that their test-passing patches leave codebases harder to maintain. We identify one concrete source: deletion avoidance, the systematic tendency to retain code that an intended edit requires removing. Across the five leading models on the official SWE-bench Verified leaderboard, deletion recall against the developer patch reaches at most 71.7\% even on tasks all five solve, and models reach the right file for over 92\% of required deletions but cut the exact line in under 52\% of cases. Instead, 29.0\% of passing patches wrap the targeted code in a guard or fallback, a pattern we call Guard-and-Go. Such patches pass because the original tests rarely check removal: when we retrofit 34 Verified tasks with tests that fail if the targeted code remains, four frontier models spanning closed and open weights fall from 63.2\% to 41.9\%. Because real repairs mix removal with addition, we curate CanItDelete, a benchmark of 200 tasks mined from real commits whose entire required edit is deletion. Even with the addition work gone, the best model still fails one task in five, and smaller open models fall to 18.0\%. We then ablate GPT-5.6 Sol under four cumulative prompts; success moves little until we supply the exact lines, which nearly eliminate incomplete deletion yet raise success only to 80.5\% because the model then deletes beyond the spans or adds code instead. Finally, through a pilot study we show one potential fix: teaching deletion during post-training reduces deletion avoidance and improves broader code-editing performance, suggesting the behavior is undertrained rather than beyond reach.
\end{abstract}

\section{Introduction}
\label{sec:intro}
\begin{figure}[t]
    \centering
    \includegraphics[width=0.9\columnwidth]{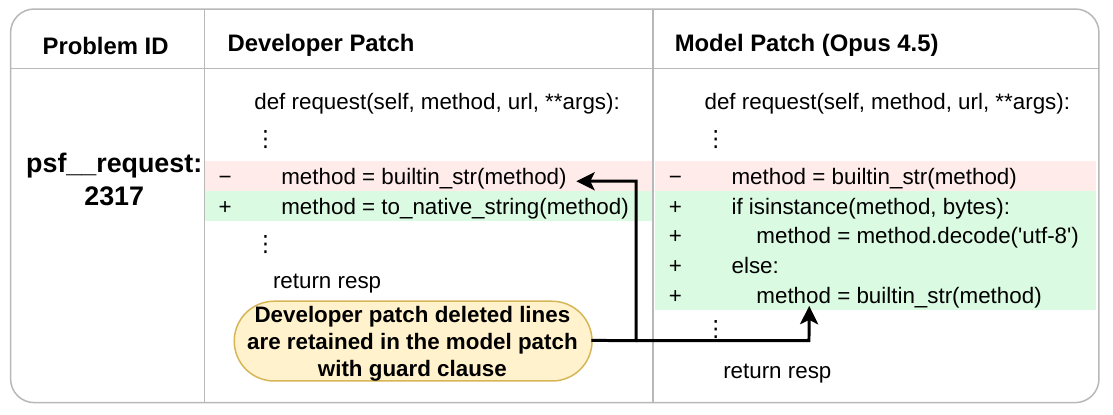}
    \caption {An example of deletion avoidance in a solved SWE-bench Verified task. The developer patch removes a statement, while the model retains it under an `else' guard.}
    \label{fig:deletion-avoidance-example}
\end{figure}
Frontier LLMs now sit at the core of coding agents that resolve issues, review code, and open pull requests with little supervision. Whether what they produce is merge-ready, meaning a maintainer would take it into the project, is a separate question from whether it runs. In a large GitHub corpus, 46.4\% of agent-authored fixes are rejected \cite{abujadallah2026understanding,alam2026unmerged}, and several projects decline AI-generated contributions outright \cite{qemu2025provenance}. Maintainers reviewing 296 already-passing agent pull requests across three SWE-bench Verified repositories merged them at a rate 24 percentage points below the benchmark score, citing verbosity and departure from repository conventions \cite{whitfill2026merge}.  %Merge readiness is now a benchmark target of its own. On a benchmark that grades scope and conformance alongside correctness, the strongest model reaches 13.4\% on the hardest tier \cite{cognition2026frontiercode}.

% FIX: "The same shift appears in aggregate commit data as a decline in
%      removal" -- "the same shift" had no antecedent; P1 names no shift.
Commit statistics point the same way. Across 623 million analyzed changes, edits that take out or update code older than twelve months fell 74\% after 2023, while error-masking constructs rose 47\% \cite{gitclear2026}. One reason agent patches read as bloated is that models leave code in place that the change was meant to remove. Reviewers of agent-written pull requests routinely delete generated methods, which they must read in full before deciding \cite{watanabe2026cut}. %The imprecision reaches single-bug debugging, where frontier models pass more than 76\% of unit tests while fewer than 45\% of their edits are necessary, a gap that instructions to change as little as possible do not close \cite{zhu2026precise}.

Figure~\ref{fig:deletion-avoidance-example} shows the behavior directly: a developer retires an obsolete assignment, while a model given the same issue keeps the assignment and routes execution around it conditionally. Both patches pass the same tests, and both are recorded as resolved. We call this \emph{deletion avoidance}, the systematic tendency to retain code that an intended edit requires removing, and we use the term for observable patch behavior rather than for internal intent.

% REWRITTEN: opens on the result. The scaffold, the five-model list, and the
%      inclusion criteria moved to Section 2, where they belong; the design is
%      now one clause. Section 3's retrofit closes the paragraph on the causal
%      link rather than starting over.
\paragraph{Deletion avoidance in the wild.}
Across five leading submissions to SWE-bench Verified, patches leave between a quarter and a third of the developer's deletions in place even on tasks that all five solve. The models are not failing to find the code: they edit the file containing a required deletion for over 92\% of target lines and the enclosing function, class, or module for 68.1\% to 74.4\%, yet remove the exact line in only 44.6\% to 51.6\% of cases. What they write instead follows one dominant strategy, in which the model keeps the logic the developer removed and adds a condition or bypass around it. We call this \emph{Guard-and-Go}, and it accounts for 29.0\% of all passing patches. Such patches pass because the original tests rarely check removal. When we retrofit 34 deletion-heavy Verified tasks with checks that fail when a validated deletion target remains, resolution across four frontier models falls 21.3 percentage points, and 29 of 86 previously passing attempts fail.

% MERGED: the benchmark and the ladder were two paragraphs restating their own
%      setups. Now one, running gap -> instrument -> result -> ladder -> result.
% \paragraph{\textsc{CanItDelete}: deletion in isolation.}
% None of this attributes the failure to deletion itself: nothing in a SWE-bench task asks for a deletion, localization is a difficulty on the benchmark in its own right \cite{alawad2026loc2repair,sepidband2026faultcontext}, and developer patches mix removal with additive work. \textsc{CanItDelete} removes those confounds, with 200 tasks mined from real commits whose entire required edit is a deletion, posed over the full pre-edit file and scored by a deterministic occurrence-aware evaluator. Across twelve models spanning frontier and open weights, success runs from 79.0\% for Claude Opus 4.8 down to 18.0\% for Qwen3-30B, and incomplete deletion is the dominant failure for ten of the twelve.

% A diagnostic ladder then adds an explicit deletion instruction, a region pointer, and finally the exact spans to remove. Even handed the spans, GPT-5.6 Sol fails 17.9\% of tasks: they cut its incomplete deletions from 18 to 6, yet success rises only from 73.4\% to 82.1\% because the failures that remain delete past the boundary or add code instead. Deletion therefore fails on both sides, too little when the boundary is uncertain and too much once it is given, so what the models lack is control rather than capability.

\paragraph{\textsc{CanItDelete}: deletion in isolation.}
None of this attributes the failure to deletion itself: nothing in a SWE-bench task asks for a deletion, localization is hard on the benchmark in its own right \cite{alawad2026loc2repair,sepidband2026faultcontext}, and developer patches mix removal with additive work. \textsc{CanItDelete} removes those confounds: 200 tasks mined from real commits whose entire required edit is deletion, posed over the full pre-edit file and scored by a deterministic occurrence-aware evaluator. Across twelve models spanning frontier and open weights, success runs from 79.0\% down to 18.0\%, and incomplete deletion dominates failures for ten of the twelve. A diagnostic ladder then adds an explicit deletion instruction, a region pointer, and finally the exact spans to remove; only the spans help every model. Even handed them, GPT-5.6 Sol still fails 19.5\% of tasks, now mostly by deleting past the boundary or adding code rather than by retaining the target. Deletion fails on both sides: too little when the boundary is uncertain, too much once it is given. What models lack is control, not capability.

% REDUCED to two lines. The old paragraph opened on "If deletion is
%      underrepresented in post-training rather than beyond reach...", the
%      conclusion of an argument this introduction no longer makes, so the
%      premise arrived from nowhere. "Learnable rather than absent" carries it.
Finally, deletion is learnable rather than absent. Adding 12.8k deletion examples to a 7B model's code post-training mixture, 0.7\% of its tokens, cuts deletion avoidance on \textsc{CanItDelete} by 13.9 percentage points, and the gain transfers to benchmarks the data never targeted, raising SWE-bench Verified by 5.3 points and CanItEdit by 1.4. Deletion may therefore be underrepresented in code post-training rather than beyond the models' reach. We report this as a proof of concept at a single scale.

\paragraph{Contributions.}
We define, characterize, and measure deletion avoidance in real repository repairs, identifying \emph{Guard-and-Go} as its dominant form and showing how much reported resolution survives a deletion-sensitive check. We release \textsc{CanItDelete}, a deletion-only benchmark with a deterministic occurrence-aware evaluator and a diagnostic ladder that localizes where deletion fails. All measurement and replication code, the benchmark, and the prompts are in the supplementary material, withheld from public release for anonymity and available on acceptance.

%\input{Prestudy}
%\input{Experiment}
%\section{Results}\label{sec:results}

\label{sec:rq1-deletion-hesitancy}
\providecommand{\NEWREF}[1]{\ref{#1}}

\section{Do LLMs Avoid Deleting Code in Practice?}
\label{sec:wild}

SWE-bench Verified measures whether a patch resolves an issue, but not how the patch transforms the code. This distinction matters when the developer repair is subtractive: a model may remove the same obsolete logic, or preserve it behind newly added control flow while still satisfying the tested behavior. We ask whether models systematically diverge from developer repairs in this direction, and what they write instead of the code developers remove.

We treat the developer patch as a behavioral reference rather than the unique correct solution. A model that retains a reference deletion may have a valid alternative repair, so a mismatch alone does not establish an error. Instead, we look for recurring evidence across models and tasks: whether models consistently delete less than developers, whether the gap remains after they reach the relevant code, and whether retained logic takes a common structural form.

\subsection{Study Design}

\noindent\textbf{Patches.}
We analyze the most recent OpenHands-based SWE-bench Verified leaderboard submission for each of five models: GLM-4.6, GPT-5, Kimi-K2, Opus-4.5, and Salesforce SAGE. Scaffolds differ in how they call tools and localize code, and those differences can shape editing behavior independently of the underlying model. We therefore hold the scaffold fixed to reduce this source of variation. Table~\ref{tab:model-selection} in Appendix~\ref{app:models} lists the submissions, their dates, and the selection procedure.

\noindent\textbf{Tasks.}
We call the human-authored patch distributed with each SWE-bench task the \emph{developer patch}. We retain a task when this patch removes at least one line from a non-test Python file without restoring that line inside the same function, class, or module; 377 of the 500 Verified tasks satisfy this criterion. We analyze the 254 tasks with unanimous outcomes: 197 that all five models solve and 57 that all five fail, setting aside the remaining 123. By restricting the comparison to unanimous outcomes, we hold both task and outcome constant across models. We read the contrast between solved and failed tasks descriptively, since tasks that no model solves may be harder for reasons unrelated to deletion.

\noindent\textbf{Reference deletions.}
A \emph{reference deletion} is a source location removed by the developer patch from a non-test Python file. We identify these locations by applying both patches to the base-commit. We discard a removed line that reappears among the developer patch's additions in the same enclosing function, class, or module, since this represents movement rather than removal. A model matches a reference deletion only by deleting the same file and location; identical text removed elsewhere does not count. We inspect every non-test Python file the model touches, including files the developer patch leaves untouched, so that we record deletions the developer did not make rather than leaving them unobserved.

\noindent\textbf{Metrics.}
Let $G_t$ be the set of reference deletions for task $t$ and $M_{t,m}$ the locations deleted by model $m$. Deletion recall is the share of the developer's deletions that the model also performs: $R_{t,m} = \frac{|G_t \cap M_{t,m}|}{|G_t|}.$
We macro-average recall across tasks so that a one-line deletion and a twenty-line deletion receive equal weight. Because recall measures how many of the developer's deletions a model reproduces, we lead with it. We also compute deletion precision, the share of model deletions that match reference locations, and report it in Table~\ref{tab:rq1-deletion-alignment} in Appendix~\ref{deletion-recall-precision-tables}. Per-task recall is bounded and skewed, so we compare solved and failed tasks with a two-sided Mann--Whitney $U$ test and report Cliff's $\delta$ with Holm-adjusted $p$-values (Table~\ref{tab:all-pass-fail-deletion-recall} in Appendix~\ref{statistical-test-for-pre-recall}).
% \NEWREF above: label your A.3 recall table as tab:recall-stats
Finally, to separate search from execution, we check each reference deletion at three nested levels: whether the model patch modifies its \emph{file}, whether it modifies at least one line within its enclosing \emph{scope}, meaning the containing function, class, or module, and whether it removes the exact \emph{line}.

\subsection{Passing Patches Still Leave Developer-Removed Code in Place}\label{sec:leaving-developer-deleted-code}

\textbf{Patches scored as correct by SWE-bench Verified still leave 28.3\% to 34.8\% of an average task's reference deletions in place.}
Across the 197 solved tasks, mean deletion recall ranges from 65.2\% for Kimi-K2 to 71.7\% for Opus-4.5 (Table~\ref{tab:deletion-recall-outcomes}). Passing the tests therefore narrows the divergence from the developer's deletions without eliminating it.

A model cannot delete code it never reaches, so we next ask whether localization accounts for this divergence. On these same 197 tasks, models modify the file containing the deletion for 92.5\% to 94.4\% of reference deletions and its enclosing scope, the containing function, class, or module, for 68.1\% to 74.4\%, yet they remove the exact line in only 44.6\% to 51.6\% of cases (Figure~\ref{fig:deletion-localization-elbow} in Appendix~\ref{elbow-curve}). Localization explains the decline from file to scope. It does not explain the further 21.4 to 27.8 percentage-point decline that occurs once the model has already modified the enclosing scope. Modifying a scope does not establish that the model inspected every line within it, but we cannot attribute the residual divergence to editing the wrong file.

The divergence is sharper still where models fail. Recall falls to between 19.8\% and 30.4\% on the 57 tasks that all five fail (Table~\ref{tab:deletion-recall-outcomes}), and the two task sets separate at Holm-adjusted $p < 10^{-8}$ for every model, with Cliff's $\delta$ between 0.485 and 0.543, a large effect throughout. Deletion precision behaves similarly (Table~\ref{tab:all-pass-fail-deletion-precision} in Appendix~\ref{statistical-test-for-pre-recall}). On solved and failed tasks alike, every model deletes less than the developer.

\begin{finding}
\noindent\textbf{Finding 1.}
Even on tasks that all five models solve, the models' patches leave 28.3\% to 34.8\% of an average task's reference deletions in place. Models edit the file containing more than 92\% of those deletions, yet remove the exact source line in only 44.6\% to 51.6\% of cases, so coarse localization explains only part of the gap.
\end{finding}

\subsection{Models Substitute Added Control Flow for Removal}

Low deletion recall may still reflect valid alternative implementations, so we examine what models write in place of the logic the developer removed. An LLM-based classifier, MiniMax-M2.7, assigns each task--model pair one of three labels. \emph{Delete-and-Replace} removes or replaces most of the developer-removed logic. \emph{Guard-and-Go} retains that logic and introduces a condition or bypass around it. A \emph{non-reference alternative} follows neither pattern and addresses the issue elsewhere. The classifier receives the issue, both patches, and precomputed deletion features, and it must support its label with lines from the supplied diffs; we reject any label whose cited evidence is absent. Because this classification does not require a solved--failed contrast, we run it over every task--model pair available in the five official leaderboard submissions rather than the 254 unanimous tasks, giving 2,487 candidate pairs.
We exclude 129 pairs whose developer patch we could not parse, leaving 2,358. 
Appendix~\ref{app:grounded-theory} reports the supplied features, the validation procedure, and the complete label definitions. The prompts are available in the replication package.

\textbf{Among the 1,703 passing pairs, 494 (29.0\%) retain the logic the developer removed and route execution around it} (Table~\ref{tab:strategy}).\footnote{Guard-and-Go accounts for 29.0\% of passing pairs and, coincidentally, 29.0\% of all 2,358 labelled pairs.} Guard-and-Go passes in 72.2\% of cases, below Delete-and-Replace at 85.2\% but well above non-reference alternatives at 39.6\%, and 190 of the 655 failing pairs follow the same strategy. The same substitution therefore recurs in patches the benchmark accepts as well as in patches it rejects.%, which motivates the removal-aware criterion we introduce in Section~\NEWREF{sec:retrofit}.% \NEWREF: label your Section 3 as sec:retrofit

We open-code the pairs, revise the categories after each round until they saturate, and consolidate the result into ten structural forms. A closed coder then applies them to all 684 Guard-and-Go pairs, assigning a form to 550 and declining on the remaining 134; Appendix~\ref{app:grounded-theory} gives the procedure and the full definitions. \emph{Retained Path as Live Fallback} accounts for 221 of the 550 typed pairs (40.2\%), more than the next three forms combined. In this form, the added guard handles the reported case while the developer-removed logic remains the default path for every other input.

Such patches preserve a developer-removed path in executable form, so a reader must judge whether both the original and the newly introduced paths remain necessary. Maintainers report the same burden when reviewing agent-authored patches: they delete generated code that they must first read \cite{watanabe2026cut}, and they cite verbosity and departure from repository conventions when rejecting patches that had passed the automated grader \cite{whitfill2026merge}.

Avoiding a deletion tends to enlarge the patch. Passing Guard-and-Go patches exceed their developer counterpart in 61.1\% of pairs, with a median size ratio of 1.67$\times$. The variation across models is substantial: 97.8\% of GLM-4.6 and 81.5\% of Kimi-K2 Guard-and-Go patches are larger, against 33.0\% for Opus-4.5 (Table~\ref{tab:guard-and-go-size} in Appendix~\ref{patch-size}). Patch size alone does not show that the added code is unnecessary. Read alongside the strategy labels, however, it shows that models often turn a subtractive developer repair into a larger patch that retains the original logic and adds a control path.

\begin{finding}
\noindent\textbf{Finding 2.}
Models substitute added control flow for removal, including in patches that SWE-bench Verified marks resolved. Guard-and-Go accounts for 29.0\% of passing patches, its dominant form retains the developer-removed logic as an executable fallback, and they exceed the developer patch in size in 61.1\% of pairs.
\end{finding}

% =====================================================================
% TABLE 1: deletion recall by outcome.
% =====================================================================
\begin{table}[t]
\centering
\footnotesize
\setlength{\tabcolsep}{4.2pt}
\renewcommand{\arraystretch}{1.04}
\begin{tabular}{@{}lrrr@{}}
\toprule
Model & Failed ($n{=}57$) & Solved ($n{=}197$) & $\delta$ \\
\midrule
GLM-4.6         & 24.0\% & 67.5\% & 0.532 \\
GPT-5           & 29.9\% & 68.5\% & 0.485 \\
Kimi-K2         & 19.8\% & 65.2\% & 0.543 \\
Opus-4.5        & 30.4\% & 71.7\% & 0.504 \\
Salesforce SAGE & 27.5\% & 68.0\% & 0.501 \\
\bottomrule
\end{tabular}
\caption{ Mean deletion recall on 197 tasks all five models solve and 57 all five fail. $\delta$ denotes effect size.}
\label{tab:deletion-recall-outcomes}
\end{table}

% =====================================================================
% TABLE 2: strategy only. Structural forms moved to Appendix B.
% =====================================================================
\begin{table}[t]
\centering\footnotesize
\setlength{\tabcolsep}{4pt}
\renewcommand{\arraystretch}{1.04}
\begin{tabular}{@{}lrrr@{}}
\toprule
Strategy & $n$ & Share & Pass rate \\
\midrule
Delete-and-Replace        & 1,197 & 50.8\% & 85.2\% \\
Guard-and-Go              &   684 & 29.0\% & 72.2\% \\
Non-reference alternative &   477 & 20.2\% & 39.6\% \\
\bottomrule
\end{tabular}
\caption{
Patch strategies among 2,358 classifier-labeled pairs. Share is overall frequency; pass rate is SWE-bench Verified resolution within each strategy. Table~\ref{tab:gng-types} (Appendix~\ref{app:grounded-theory}) lists the ten Guard-and-Go forms.}
\label{tab:strategy}
\end{table}

\section{Do Passing Tests Detect Missing Deletions?}
\label{sec:f2p}

\begin{table}[t]
\centering
\footnotesize
\setlength{\tabcolsep}{3.5pt}
\renewcommand{\arraystretch}{1.08}

\begin{tabular}{@{}lrrrr@{}}
\toprule
\multicolumn{2}{c}{} &
\multicolumn{2}{c}{\textbf{Passing attempts}} &
\\
\cmidrule(lr){3-4}
\textbf{Model} &
\textbf{Tasks} &
\begin{tabular}[c]{@{}c@{}}
\textbf{Original}\\
\textbf{suite}
\end{tabular} &
\begin{tabular}[c]{@{}c@{}}
\textbf{With deletion}\\
\textbf{check}
\end{tabular} &
\begin{tabular}[c]{@{}c@{}}
\textbf{Drop}\\
\textbf{(pp)}
\end{tabular} \\
\midrule
GPT-5.6 Sol     &  34 & 21 (61.8\%) & 15 (44.1\%) & 17.6 \\
Opus 4.8        &  34 & 21 (61.8\%) & 14 (41.2\%) & 20.6 \\
GLM-5.2         &  34 & 26 (76.5\%) & 18 (52.9\%) & 23.5 \\
DeepSeek-V4-Pro &  34 & 18 (52.9\%) & 10 (29.4\%) & 23.5 \\
\midrule
\textbf{Overall} &
\textbf{136} &
\textbf{86 (63.2\%)} &
\textbf{57 (41.9\%)} &
\textbf{21.3} \\
\bottomrule
\end{tabular}

\caption{Attempts passing the original suite and, among them, the deletion-sensitive check on 34 tasks per model. Drop is in percentage points (pp); each task is about 2.9 pp.}
\label{tab:deletion-f2p-impact}
\end{table}

Section \ref{sec:wild} shows that models repeatedly retain developer-removed logic behind added guards and bypasses, including in patches that SWE-bench Verified records as resolved. Comparing model patches with the developer patch cannot determine how to interpret this divergence. A retained target may belong to an adequate alternative repair, or the original tests may not require its removal. We thus modify the evaluation criterion rather than the model, retrofitting Verified tasks with checks that fail when a validated deletion target remains and measuring how much reported resolution survives.

Among the 69 Verified tasks in which deletions constitute at least a quarter of the developer patch, we identify a substantive deletion target using an AST-aware procedure that prioritizes removed conditions, control-flow statements, and complete blocks. We then construct a source-level \emph{deletion-sensitive check} requiring the target to be absent from its enclosing scope, and retain a task only when the check fails on the base revision and passes after applying the developer patch. This leaves 34 tasks with validated deletion targets and a demonstrated repair that removes them (see the supplementary material at Appendix~\ref{app:deletion-f2p}).

Rather than reusing the submissions we analyze in Section~\ref{sec:wild}, we generate patches with four frontier models spanning closed and open weights: GPT-5.6 Sol, Opus 4.8, GLM-5.2, and DeepSeek-V4-Pro. Those submissions all predate December 2025, so applying a deletion-sensitive check to them would leave open whether current systems still exhibit the behavior \cite{swebench2025policy,openai2026retire}. We use the same models in Section~\ref{sec:canitdelete-construction}, where we examine deletion under controlled conditions.

\textbf{Across the 136 attempts we generate, 86 (63.2\%) pass the original test suites, and 57 (41.9\%) also satisfy the deletion-sensitive check.} The absolute decline is 21.3 percentage points, and 29 of the 86 accepted attempts (33.7\%) retain the validated target. Every model declines, by 17.6 to 23.5 percentage points (Table~\ref{tab:deletion-f2p-impact}).

% Because each task and model patch remain fixed across the two criteria, the paired decline follows from strengthening the evaluation criterion rather than from changing the task or generation setting. On these deletion-heavy tasks, roughly one third of the patches accepted by the original suites retain the validated target even though a developer repair removes it.

Because each task and model patch is fixed under both criteria, the paired decline reflects stricter evaluation, not a change in the task or generation setting. On these deletion-heavy tasks, roughly one third of patches accepted by the original suites retain a validated target that the developer repair removes.

However, we derive each check from a target removed by the developer patch, and our experiment measures performance under that removal requirement rather than establishing that deletion is the only behaviorally valid repair. An alternative repair may preserve the target and still satisfy the original behavioral specification. The 34 tasks are also deletion-heavy by construction and do not represent SWE-bench Verified as a whole. The retrofit therefore identifies an evaluation gap without isolating why models fail to produce a removing repair. Full repository tasks still entangle the decision to delete with localization, boundary identification, and accompanying implementation work.

\begin{finding}
\noindent\textbf{Finding 3.}
Adding a deletion-sensitive check reduces resolution by 21.3 percentage points on deletion-heavy SWE-bench Verified tasks. Of the attempts the original suites accept, 33.7\% retain the validated target, indicating a substantial gap between passing the behavioral tests and satisfying an explicit removal requirement.

\end{finding}

\section{CanItDelete: A Diagnostic Benchmark for Deletion Avoidance}\label{sec:canitdelete-construction}
 \begin{figure*}[t]
      \centering
      \includegraphics[width=0.8\textwidth]{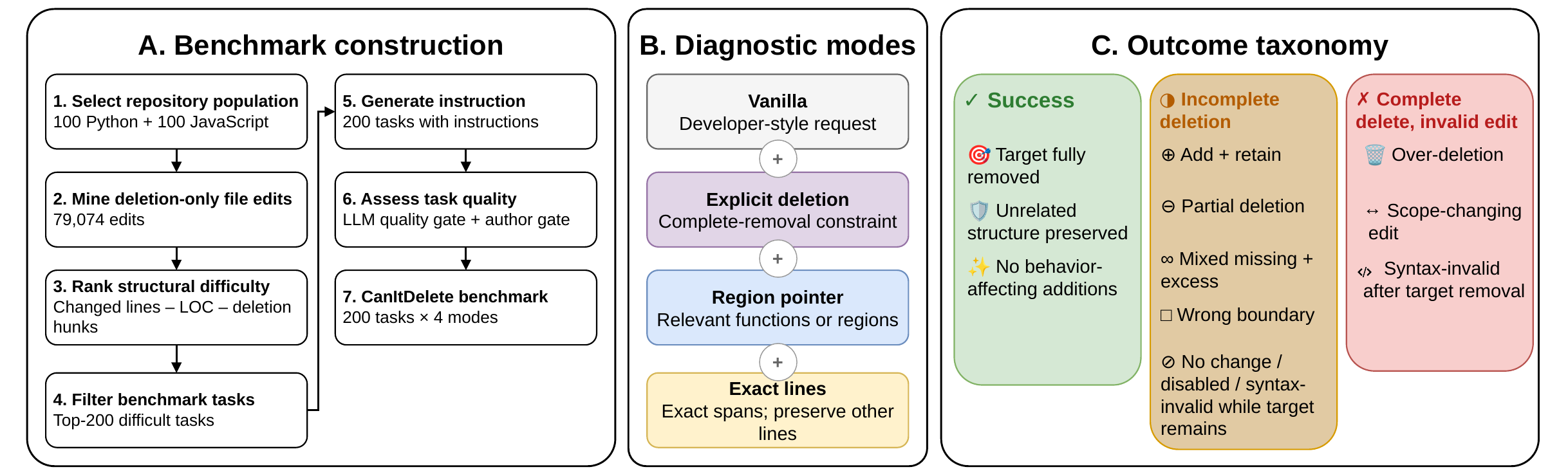}
      \caption{Overview of CanItDelete benchmark construction, cumulative diagnostic modes, and the structural outcome
      taxonomy.}
      \label{fig:overview}
  \end{figure*}
Sections~\ref{sec:wild} and~\ref{sec:f2p} establish that models leave developer-removed code in place and that behavioral test suites accept many of the patches that do. Neither study attributes the failure to deletion itself: in full repository repair, a missed removal can originate in localization, in replacement code, or in surrounding implementation work, and nothing in the task states that removal is required. We therefore build \textsc{CanItDelete}, 200 tasks mined from real commits in which deletion is the complete required transformation. Because the reference edit adds nothing, every compliant solution must perform the same removal while preserving unrelated code, and a failed output reflects the model's editing behavior rather than ambiguity about what the task demands. Figure~\ref{fig:overview} summarizes the design: the construction pipeline (A), the signal ladder, four cumulative modes that each add one localizing cue (B), and the failure taxonomy our evaluator assigns (C). The taxonomy and the ladder are what let the benchmark diagnose rather than rank. Appendix~\ref{app:canitdelete-design} details benchmark construction and validation, and Appendix~\ref{app:canitdelete-evaluator} specifies the deletion-compliance evaluator.

 \begin{figure}[t]
      \centering
      \includegraphics[width=\columnwidth]
      {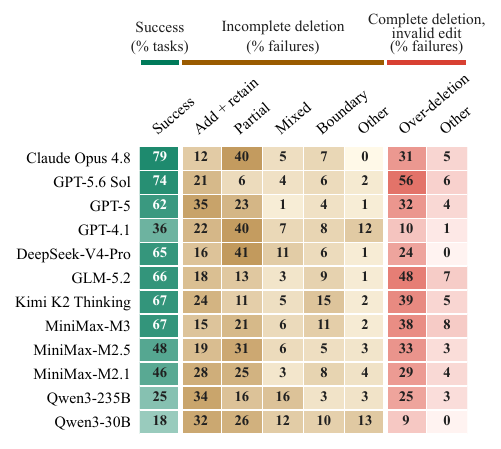}
      \caption{Vanilla-mode success and failure composition across 12 models. Success is measured over 200 tasks;
      failure mechanisms are measured over each model's failed tasks.}
      \label{fig:vanilla}
  \end{figure}

\subsection{Benchmark Design}\label{subsec:benchmark_design}
 
\paragraph{Construction.}

From the 100 most-starred active repositories in each of Python and JavaScript, we mine 79,074 file edits that delete source lines and add none. We rank them by an equal-weight index over pre-edit length, deleted lines, and deletion hunks, three change-complexity dimensions established in software-maintenance research \cite{nagappan2005relative,kamei2013large}, and keep the 200 hardest, one file per repository--commit pair (Figure~\ref{fig:overview}A). We select deliberately for stress: every task spans at least three separated deletion hunks, so the benchmark emphasizes multi-site deletion over one-line cleanup. The 200 tasks come from 35 repositories; 151 are Python, 49 JavaScript-family, and 53 modify test files. We generate instructions only after selection, using GPT-5.6 Sol to draft a short deletion request from the complete pre-edit file and reference diff, and we pass every task through an LLM rubric and an author audit, both requiring the request to cover every substantive deletion and to be locatable from the pre-edit file alone (Appendix~\ref{app:instruction-gen-task-val}). 
 
\paragraph{Evaluation.}
We score outputs with a deterministic, occurrence-aware evaluator; no LLM judges any output. An output is deletion-compliant when the complete target is absent, 
executable structure outside it is preserved, and no behavior-affecting or unrelated change is introduced. Commenting out or disabling the target does not count, and deleting one copy of a repeated line earns no credit for a different required occurrence. We partition failures by whether required code remains into \emph{incomplete deletion} and \emph{complete deletion with an invalid edit}; Figure~\ref{fig:overview}C names the mechanisms within each, from add-and-retain to over-deletion. The diagnostic ladder supplies four cumulative modes, vanilla, explicit deletion, region pointer, and exact lines, each adding one cue, so the change in a model's outcomes from one mode to the next reveals whether intent, search, or boundary knowledge was missing (Figure~\ref{fig:overview}B; definitions in Appendix~\ref{app:canitdelete-modes}).
 
\subsection{Deletion Avoidance Persists in Isolation}
 
\textbf{The best performing frontier model fails one deletion-only task in five.} Claude Opus 4.8 leads the twelve models we evaluate at 79.0\% deletion-compliant success, and GPT-5.6 Sol follows at 74.0\% (Figure~\ref{fig:vanilla}). These failures occur even though we remove the confounds identified above: we supply the complete file, we ask only for removal, and the task requires no cross-file localization and no replacement code. The avoidance we measure in the wild in Section~\ref{sec:wild} therefore persists when the model receives the file directly and deletion is the entire task, and it takes the same form. Add-and-retain, the dominant failure mechanism here, is Guard-and-Go under controlled conditions: an addition standing in for a deletion.

\textbf{Open-weight leaders trail the frontier by roughly twelve points, and the remaining open models fall much further.} The strongest of them, Kimi K2 Thinking, MiniMax-M3, GLM-5.2, and DeepSeek-V4-Pro, cluster within a narrow 65.0--67.0\% band, whereas the Qwen instruct models and the earlier MiniMax releases reach only 18.0--47.5\% and fail predominantly by leaving required code behind. The task set separates these models rather than saturating at either end: 9 of the 200 tasks are solved by all twelve models, 19 by none, and every intermediate solved-by count occurs in between.
 
\textbf{Behind these pass rates sit two distinct failure regimes.} Incomplete deletion is the majority failure for ten of twelve models (69.8\% pooled), but GPT-5.6 Sol and GLM-5.2 define a second regime: they usually remove the target and then over-delete or edit out of scope. Which regime a model falls into is not fixed by capability. Along the GPT line in Figure~\ref{fig:vanilla}, incomplete deletions fall from 114 to 20 while invalid edits after complete removal rise from 14 to 32, and Qwen shows the same exchange at lower capability, whereas MiniMax-M3 reduces both. A single pass rate would record progress here; only the mechanism decomposition shows what kind.
 
\begin{finding}
\textbf{Finding 4.} Across twelve models, deletion-only success ranges from 18.0\% to 79.0\%, and incomplete deletion accounts for 69.8\% of failures. Stronger deletion behavior does not always become compliant editing: some models replace target retention with over-deletion, revealing distinct problems of deletion completion and scope preservation.

\end{finding}
 
\subsection{Models Fail Even When Given the Exact Lines to Delete}
 
\begin{figure*}[t]
      \centering
      \includegraphics[width=0.9\textwidth]
      {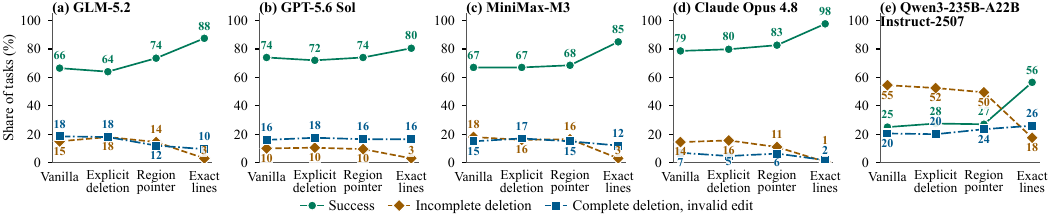}
      \caption{Diagnostic-ladder outcomes under increasingly precise deletion guidance. Four models use 200 tasks; Claude Opus 4.8 uses 173.}
      \label{fig:ladder}
  \end{figure*}

\textbf{Even given the exact lines to delete, no model is flawless, and only one comes close.} With occurrence-specific spans supplied, Claude Opus 4.8 reaches 97.7\%, while the other four ladder models finish between 56.5\% and 87.5\% (Figure~\ref{fig:ladder}). Qwen3-235B still leaves required code in 17.5\% of tasks after being told exactly what to remove, the clearest sign that deletion avoidance persists, and across the five models 1.7--26.0\% of attempts fail even after the complete target is gone because the edit strays beyond it.

\textbf{Exact spans are the only signal that moves every model.} They raise success by 6.5--31.5 points and cut incomplete deletion to 0.6--3.0\% for four of the five, whereas the cheaper signals accomplish little: an explicit no-workaround instruction shifts success by only $-2.5$ to $+2.5$ points. Region pointers change success by 0.0--7.0 points, with the largest gain for GLM-5.2. Deletion avoidance is not a misreading of intent, and search is not the main bottleneck; until we hand over the spans, what models lack is knowledge of where the deletion ends.

\textbf{Suppressing incomplete deletion exposes a second tendency to over-edit.} GPT-5.6 Sol's invalid-edit rate barely moves with exact lines, from 16.0\% to 16.5\%, and Qwen3-235B's rises from 20.5\% to 26.0\% as its retention falls. The ladder thus separates two capabilities that a single pass rate conflates, finding every required occurrence and stopping at its boundary: aggregate success cannot tell which side a model improved, or whether, like Qwen, it traded one failure for the other, while the mode deltas and the mechanism split can. Current models lack control rather than capability. We report per-model results for all four modes in Appendix~\ref{app:canitdelete-ladder-details}.

\begin{finding}
\textbf{Finding 5.} Exact deletion spans raise success by 6.5--31.5 points and nearly eliminate incomplete deletion for four of five models. Nevertheless, as many as 26.0\% of tasks still fail after complete target removal, showing that exact localization and disciplined scope preservation are distinct capabilities.
\end{finding}

\section{Can Deletion-Focused Post-Training Reduce Deletion Avoidance?}
\label{sec:mitigation}

Section~\ref{sec:canitdelete-construction} shows that models can perform deletion but do not reliably choose it or preserve its boundary. One plausible explanation combines the additive preference observed in people and language models with the action bias observed in coding agents: when a change requires deletion, models may default to acting through added code \cite{adams2021subtractive,santagata2025more,gloaguen2026fixedbench}. Neither the data nor the objective corrects this. The additive skew is present in the text models learn from \cite{winter2023more}, and evaluation based on behavioral correctness accepts Guard-and-Go, as Section~\ref{sec:f2p} showed that a third of patches accepted by the original suites retain the deletion target. We hypothesize that deletion is insufficiently reinforced during code post-training, both in whether models choose it and in where they stop.

Because deletion is a cross-cutting code-editing behavior rather than a standalone downstream task, we add deletion supervision to a general code post-training mixture rather than train a deletion-specific adapter.

\paragraph{Proof-of-concept intervention.}

We use a 7B in-house model because a larger member of the same family supports coding workflows in our industrial setting.\footnote{Model and hardware details are anonymized.} We augment its code-only post-training mixture with 12,821 deletion examples: 10,000 file-level edits and 2,821 repository-level repairs constructed using the \textsc{CanItDelete} pipeline. These contribute 112.1M tokens to the 15.9B-token mixture, approximately 0.7\% of the total. Appendix~\ref{app:sft-selection} gives the construction and rejection procedure; we exclude all \textsc{CanItDelete} evaluation problems from training.

Both checkpoints are trained under an identical recipe, six epochs with a
global batch size of 64 on 128 xPUs. The baseline is post-trained on the
15.9B-token code-only mixture; the intervention adds the deletion subset to
that same mixture and is otherwise unchanged, so the two runs differ only in
0.7\% of training tokens. CanItDelete measures the targeted behavior. The
remaining three cover the code-editing workloads the deployed model serves:
SWE-bench Verified evaluates repository-level repair from real issue reports
\cite{jimenez2024swebench,openai2024verified}, while CanItEdit and EditBench
evaluate instructed edits to existing code from a natural-language request
\cite{cassano2024canitedit,chi2025editbench}. None targets deletion, so
together they test whether the intervention transfers or introduces
regressions. We run inference three times on every benchmark and report the
mean.

\paragraph{Deletion training reduces incomplete deletion and exposes failures of scope preservation.}
\textsc{CanItDelete} success increases from 6.5\% to 13.7\%, while incomplete deletion falls from 80.4\% to 66.5\% (Table~\ref{tab:posttraining}). The 13.9 percentage-point reduction splits almost evenly: 7.2 points become compliant edits, while 6.7 become complete-but-invalid edits, and over-deletion alone rises by 6.2 points. The intervention therefore makes the model more likely to complete the required removal, but does not yet teach it where that removal should stop.

\begin{table}[t]
\centering
\scriptsize
\setlength{\tabcolsep}{2.0pt}
\renewcommand{\arraystretch}{1.05}
\begin{tabular}{@{}lrrrrrrr@{}}
\toprule
& \multicolumn{4}{c}{\textit{CanItDelete outcome}}
& \multicolumn{3}{c}{\textit{Benchmark success}} \\
\cmidrule(lr){2-5}
\cmidrule(l){6-8}
& Succ.
& Incomp.
& \shortstack{Complete\\invalid}
& \shortstack{Over-\\deletion}
& \shortstack{SWE-b.\\Verified}
& EditBench
& CanItEdit \\
\midrule
Base
&  6.5 & 80.4 & 13.1 & 10.6
& 25.40 & 39.26 & 44.30 \\
+Deletion
& 13.7 & 66.5 & 19.8 & 16.8
& 30.70 & 39.07 & 45.70 \\
\midrule
$\Delta$
& +7.2 & $-13.9$ & +6.7 & +6.2
& +5.30 & -0.19 & +1.40 \\
\bottomrule
\end{tabular}
\caption{
 7B-model performance before and after deletion-augmented post-training. Values are three-run means (\%); $\Delta$ is absolute percentage-point change. Over-deletion is a subset of complete-but-invalid edits.}
\label{tab:posttraining}
\end{table}

The gain concentrates where removal is part of the task. SWE-bench Verified rises 5.3 percentage points, while CanItEdit rises 1.40 and EditBench is unchanged at $-0.19$: 377 of the 500 Verified tasks require at least one deletion from a non-test file (Section~\ref{sec:wild}), whereas neither instructed-editing benchmark is deletion-oriented. The effect is therefore selective rather than uniform, which is what a deletion-specific mechanism predicts.

None of the studied benchmarks declines by more than 0.2 points, so the intervention clears the regression gate this pilot was built to test. We read the result as a signal rather than a solution: reducing the additive substitution did not remove the disposition to act on the code, which now surfaces as over-deletion, so deletion completion and boundary control are distinct training objectives. Whether either holds at deployment scale is left to future work.

\begin{finding}
\noindent\textbf{Finding 6. }In a single-model pilot, adding deletion supervision to a code post-training mixture reduces incomplete deletion by 13.9 percentage points. 
\end{finding}

\section{Related Work}\label{sec:related-work}
\paragraph{Additive bias and deletion behavior.}
People systematically favor additive changes over subtractive ones \cite{adams2021subtractive}, a skew that English corpus statistics reproduce \cite{winter2023more} and that LLMs inherit on controlled tasks \cite{santagata2025more}. Coding agents show a related tendency, acting when inaction would be correct \cite{gloaguen2026fixedbench}. LLM patches can also err in the opposite direction, removing unrelated code and breaking working behavior \cite{contributors2026}. These studies all work in synthetic, no-op, or adjacent editing settings, and none asks whether models retain developer-removed code in real repository repairs.
 
% CUT: "Most code-editing evaluations treat a patch as correct when it passes
%      the benchmark tests" -- the premise is Section 3's opening.
% CUT: the Ma et al. library example, folded into the specification-gaming
%      citation rather than spelled out.
\paragraph{Validity of test-based code-editing evaluation.}
Passing a benchmark's tests does not establish that the requested change was implemented correctly. Audits of SWE-bench report weak tests, mislabeled passes, and flawed task specifications \cite{aleithan2024swebenchplus,yu2025utboost,openai2026retire}. Coverage is narrow as well: 56\% of EDIT-Bench tests exercise only the edited region \cite{editverify2026}, which leaves room for specification gaming \cite{krakovna2020specification,ma2026buildingtotest}. Deletion-only Kali patches make the complementary point, passing tests without repairing the defect \cite{ginelli2022coderemoval}. None of this work isolates deletion as the target behavior.
 
% CUT: SWE-bench's 2,294 issues and the 500-instance description. The reader
%      has been reading about Verified for six pages, and 2,294 is used nowhere.
% FIX: "deletion-boundary control" -> scope preservation (Findings 4 and 6)
\paragraph{Benchmarks for code editing and repair.}
SWE-bench and its Verified subset evaluate repository-level repair from real issue reports \cite{jimenez2024swebench,openai2024verified}, and instructed-editing benchmarks evaluate modifications requested in natural language \cite{cassano2024canitedit,chi2025editbench,guo2025codeeditorbench}. Prior work also shows that supplying localization information improves repair \cite{alawad2026loc2repair,sepidband2026faultcontext}. In every case the reference patch combines addition, modification, and removal, so none isolates an LLM's ability to execute a purely subtractive edit. \textsc{CanItDelete} makes deletion the required behavior and adds a diagnostic ladder that separates failures of intent, localization, and scope preservation.

\section{Conclusion and Limitations}
\label{sec:conclusion}

% HALVED per page budget. CUT: the definition restatement, the
%      maintainer-reading-burden clause (Section 2.3 carries it), the
%      transfer specifics, and "we measure neither effect".
Deletion avoidance recurs across current code models: patches that SWE-bench Verified marks resolved retain a quarter to a third of the developer's deletions, substituting added control flow for removal, and pass because the tests rarely check it, so resolution rates overstate merge-ready behavior. The gap persists when deletion is the entire task, and exact spans only trade retention for over-deletion: models lack control over removal rather than the capability. Modest deletion supervision reduces the behavior and improves repository-level repair, so the deficit appears undertrained rather than intrinsic.

These findings are bounded in scope, construction, and scale. The in-the-wild analysis rests on submitted SWE-bench Verified patches with uncontrolled decoding settings, and the deletion-sensitive check covers 34 deletion-heavy tasks. \textsc{CanItDelete} instructions are drafted with GPT-5.6 Sol, itself an evaluated model, from most-starred repositories whose post-edit files may appear in training data. The pilot trains one 7B model and reports three-run means without variance; whether deletion supervision holds at deployment scale or across other languages remains open.

\bibliography{main}

\clearpage
\appendix

\section{Analyzing Deletion Avoidance Quantatitively}
\subsection{Model Choices for Empirical Analysis}\label{app:models}

To control for variation introduced by agent scaffolding, we selected models from the official SWE-bench Verified submissions repository using a consistent procedure. We first identified submissions using the OpenHands scaffold across both open-weight and proprietary model families. We then excluded submissions older than one year, using July 1, 2025, as the cut-off date. Finally, for each eligible model family, we selected its latest OpenHands-based submission. This procedure provides coverage across diverse model families while holding the agent scaffold constant and ensuring that the evaluated submissions are recent. Selected model summary is in Table~\ref{tab:model-selection}

% \begin{table}[h]
% \centering
% \footnotesize
% \setlength{\tabcolsep}{2.5pt}
% \renewcommand{\arraystretch}{1.15}

% \caption{Model submissions selected from the official SWE-bench Verified
% repository. For each model family, we retained the latest submission using
% the OpenHands scaffold. The submission date is extracted from the leading
% \texttt{YYYYMMDD} component of the repository directory name.}
% \label{tab:model-selection}

% \begin{tabular}{
%     >{\raggedright\arraybackslash}p{0.9cm}
%     >{\raggedright\arraybackslash}p{1.7cm}
%     >{\raggedright\arraybackslash}p{3.2cm}
%     >{\raggedright\arraybackslash}p{1.4cm}
% }
% \toprule
% \textbf{Family} &
% \parbox[t]{1.7cm}{\raggedright\textbf{Selected model}} &
% \parbox[t]{3.2cm}{\raggedright\textbf{Submission directory}} &
% \parbox[t]{1.4cm}{\raggedright\textbf{Submission date}} \\
% \midrule

% GLM &
% GLM-4.6 &
% \parbox[t]{3.2cm}{\ttfamily
% 20250930\_zai\_\\
% glm-4.6} &
% \parbox[t]{1.4cm}{September 30,\\2025} \\

% GPT &
% GPT-5 &
% \parbox[t]{3.2cm}{\ttfamily
% 20250807\_\\
% openhands\_gpt5} &
% \parbox[t]{1.4cm}{August 7,\\2025} \\

% Kimi &
% Kimi K2 &
% \parbox[t]{3.2cm}{\ttfamily
% 20250716\_\\
% openhands\_kimi\_k2} &
% \parbox[t]{1.4cm}{July 16,\\2025} \\

% Claude &
% \parbox[t]{1.7cm}{Claude\\Opus 4.5} &
% \parbox[t]{3.2cm}{\ttfamily
% 20251127\_\\
% openhands\_\\
% claude-opus-4-5} &
% \parbox[t]{1.4cm}{November 27,\\2025} \\

% SAGE &
% \parbox[t]{1.7cm}{Salesforce\\SAGE} &
% \parbox[t]{3.2cm}{\ttfamily
% 20251103\_\\
% SalesforceAIResearch\_\\
% SAGE\_OpenHands} &
% \parbox[t]{1.4cm}{November 3,\\2025} \\

% \bottomrule
% \end{tabular}
% \end{table}
\begin{table}[h]
\centering
\small
\setlength{\tabcolsep}{2.5pt}
\renewcommand{\arraystretch}{1.15}

\begin{tabular}{
    >{\raggedright\arraybackslash}p{0.9cm}
    >{\raggedright\arraybackslash}p{1.7cm}
    >{\raggedright\arraybackslash}p{3.2cm}
    >{\raggedright\arraybackslash}p{1.4cm}
}
\toprule
\textbf{Family} &
\parbox[t]{1.7cm}{\raggedright\textbf{Selected model}} &
\parbox[t]{3.2cm}{\raggedright\textbf{Submission directory}} &
\parbox[t]{1.4cm}{\raggedright\textbf{Submission date}} \\
\midrule

GLM &
GLM-4.6 &
\parbox[t]{3.2cm}{\ttfamily
20250930\_zai\_\\
glm-4.6} &
\parbox[t]{1.4cm}{September 30,\\2025} \\

GPT &
GPT-5 &
\parbox[t]{3.2cm}{\ttfamily
20250807\_\\
openhands\_gpt5} &
\parbox[t]{1.4cm}{August 7,\\2025} \\

Kimi &
Kimi K2 &
\parbox[t]{3.2cm}{\ttfamily
20250716\_\\
openhands\_kimi\_k2} &
\parbox[t]{1.4cm}{July 16,\\2025} \\

Claude &
\parbox[t]{1.7cm}{Claude\\Opus 4.5} &
\parbox[t]{3.2cm}{\ttfamily
20251127\_\\
openhands\_\\
claude-opus-4-5} &
\parbox[t]{1.4cm}{November 27,\\2025} \\

SAGE &
\parbox[t]{1.7cm}{Salesforce\\SAGE} &
\parbox[t]{3.2cm}{\ttfamily
20251103\_\\
SalesforceAIResearch\_\\
SAGE\_OpenHands} &
\parbox[t]{1.4cm}{November 3,\\2025} \\

\bottomrule
\end{tabular}

\caption{Model submissions selected from the official SWE-bench Verified
repository. For each model family, we retained the latest submission using
the OpenHands scaffold. The submission date is extracted from the leading
\texttt{YYYYMMDD} component of the repository directory name.}
\label{tab:model-selection}
\end{table}

\subsection{Deletion Precision and Recall between Developer Patch and Model Patch}\label{deletion-recall-precision-tables}

\begin{table}[!htbp]
\centering
\small
\setlength{\tabcolsep}{3.5pt}
\renewcommand{\arraystretch}{1.05}

\begin{tabular}{lrrrr}
\toprule
& \multicolumn{2}{c}{\begin{tabular}[c]{@{}c@{}}
\textbf{Failed by all five}\\
($n=57$)
\end{tabular}}
& \multicolumn{2}{c}{\begin{tabular}[c]{@{}c@{}}
\textbf{Solved by all five}\\
($n=197$)
\end{tabular}} \\
\cmidrule(lr){2-3}
\cmidrule(lr){4-5}
\textbf{Model} &
\textbf{Precision} &
\textbf{Recall} &
\textbf{Precision} &
\textbf{Recall} \\
\midrule
GLM-4.6         & 34.6\% & 24.0\% & 70.7\% & 67.5\% \\
GPT-5           & 35.6\% & 29.9\% & 69.3\% & 68.5\% \\
Kimi-K2         & 35.0\% & 19.8\% & 70.6\% & 65.2\% \\
Opus-4.5        & 47.0\% & 30.4\% & 74.3\% & 71.7\% \\
Salesforce SAGE & 36.7\% & 27.5\% & 66.2\% & 68.0\% \\
\bottomrule
\end{tabular}

\caption{Mean per-task deletion precision and recall. A match requires the
model-generated patch to delete the same base-commit source location as the
official human patch.}
\label{tab:rq1-deletion-alignment}
\end{table}

\FloatBarrier

\subsection{Statistical Test for Precision and Recall}\label{statistical-test-for-pre-recall}

\begin{table}[!htbp]
\centering
\small
\setlength{\tabcolsep}{4pt}
\renewcommand{\arraystretch}{1.05}

\textbf{Panel A: Descriptive statistics}

\vspace{2pt}

\begin{tabular}{lrrrr}
\toprule
& \multicolumn{2}{c}{\textbf{All-passed}}
& \multicolumn{2}{c}{\textbf{All-failed}} \\
\cmidrule(lr){2-3}
\cmidrule(lr){4-5}
\textbf{Model} &
\textbf{Median} &
\textbf{IQR} &
\textbf{Median} &
\textbf{IQR} \\
\midrule
Salesforce SAGE & 1.00 & 0.75 & 0.07 & 0.50 \\
GLM-4.6         & 1.00 & 0.75 & 0.00 & 0.33 \\
GPT-5           & 1.00 & 0.71 & 0.00 & 0.50 \\
Kimi-K2         & 1.00 & 0.80 & 0.00 & 0.25 \\
Opus-4.5        & 1.00 & 0.61 & 0.17 & 0.50 \\
\bottomrule
\end{tabular}

\vspace{6pt}

\textbf{Panel B: Statistical comparison}

\vspace{2pt}

\begin{tabular}{lrrrr}
\toprule
\textbf{Model} &
\textbf{\(U\)} &
\textbf{\begin{tabular}[c]{@{}r@{}}
Holm-adjusted\\\(p\)
\end{tabular}} &
\textbf{Cliff's \(\delta\)} &
\textbf{Effect} \\
\midrule
Salesforce SAGE & 8,426.50 & \(1.29 \times 10^{-9}\)  & 0.50 & Large \\
GLM-4.6         & 8,599.50 & \(2.41 \times 10^{-10}\) & 0.53 & Large \\
GPT-5           & 8,338.00 & \(1.99 \times 10^{-9}\)  & 0.49 & Large \\
Kimi-K2         & 8,662.00 & \(1.59 \times 10^{-10}\) & 0.54 & Large \\
Opus-4.5        & 8,444.50 & \(6.51 \times 10^{-10}\) & 0.50 & Large \\
\bottomrule
\end{tabular}

\caption{Comparison of deletion recall between all-passed (\(n=197\)) and
all-failed (\(n=57\)) eligible tasks for each evaluated model. Panel A reports
the median and interquartile range (IQR). Panel B reports the Mann--Whitney
\(U\) test with Holm-adjusted \(p\)-values and Cliff's \(\delta\) effect size.}
\label{tab:all-pass-fail-deletion-recall}
\end{table}

\begin{table}[!htbp]
\centering
\small
\setlength{\tabcolsep}{4pt}
\renewcommand{\arraystretch}{1.05}

\textbf{Panel A: Descriptive statistics}

\vspace{2pt}

\begin{tabular}{lrrrr}
\toprule
& \multicolumn{2}{c}{\textbf{All-passed}}
& \multicolumn{2}{c}{\textbf{All-failed}} \\
\cmidrule(lr){2-3}
\cmidrule(lr){4-5}
\textbf{Model} &
\textbf{Median} &
\textbf{IQR} &
\textbf{Median} &
\textbf{IQR} \\
\midrule
Salesforce SAGE & 1.000 & 0.684 & 0.038 & 0.929 \\
GLM-4.6         & 1.000 & 0.667 & 0.000 & 1.000 \\
GPT-5           & 1.000 & 0.684 & 0.000 & 1.000 \\
Kimi-K2         & 1.000 & 0.667 & 0.000 & 1.000 \\
Opus-4.5        & 1.000 & 0.500 & 0.400 & 1.000 \\
\bottomrule
\end{tabular}

\vspace{6pt}

\textbf{Panel B: Statistical comparison}

\vspace{2pt}

\begin{tabular}{lrrrr}
\toprule
\textbf{Model} &
\textbf{\(U\)} &
\textbf{\begin{tabular}[c]{@{}r@{}}
Holm-adjusted\\
\(p\)
\end{tabular}} &
\textbf{Cliff's \(\delta\)} &
\textbf{Effect} \\
\midrule
Salesforce SAGE & 7,707.5 & \(9.34 \times 10^{-6}\) & 0.373 & Medium \\
GLM-4.6         & 8,014.0 & \(3.05 \times 10^{-7}\) & 0.427 & Medium \\
GPT-5           & 7,861.0 & \(1.49 \times 10^{-6}\) & 0.400 & Medium \\
Kimi-K2         & 7,867.5 & \(9.67 \times 10^{-7}\) & 0.401 & Medium \\
Opus-4.5        & 7,504.0 & \(1.01 \times 10^{-5}\) & 0.337 & Medium \\
\bottomrule
\end{tabular}

\caption{Comparison of deletion precision between all-passed (\(n=197\)) and
all-failed (\(n=57\)) eligible tasks for each evaluated model. Panel A reports
the median and interquartile range (IQR). Panel B reports the Mann--Whitney
\(U\) test with Holm-adjusted \(p\)-values and Cliff's \(\delta\) effect size.}
\label{tab:all-pass-fail-deletion-precision}
\end{table}

\FloatBarrier

\subsection{Exploring the Deletion Avoidance from Localization point of View}\label{elbow-curve}
\begin{figure}[!htbp]
  \centering
  \includegraphics[width=\columnwidth]
  {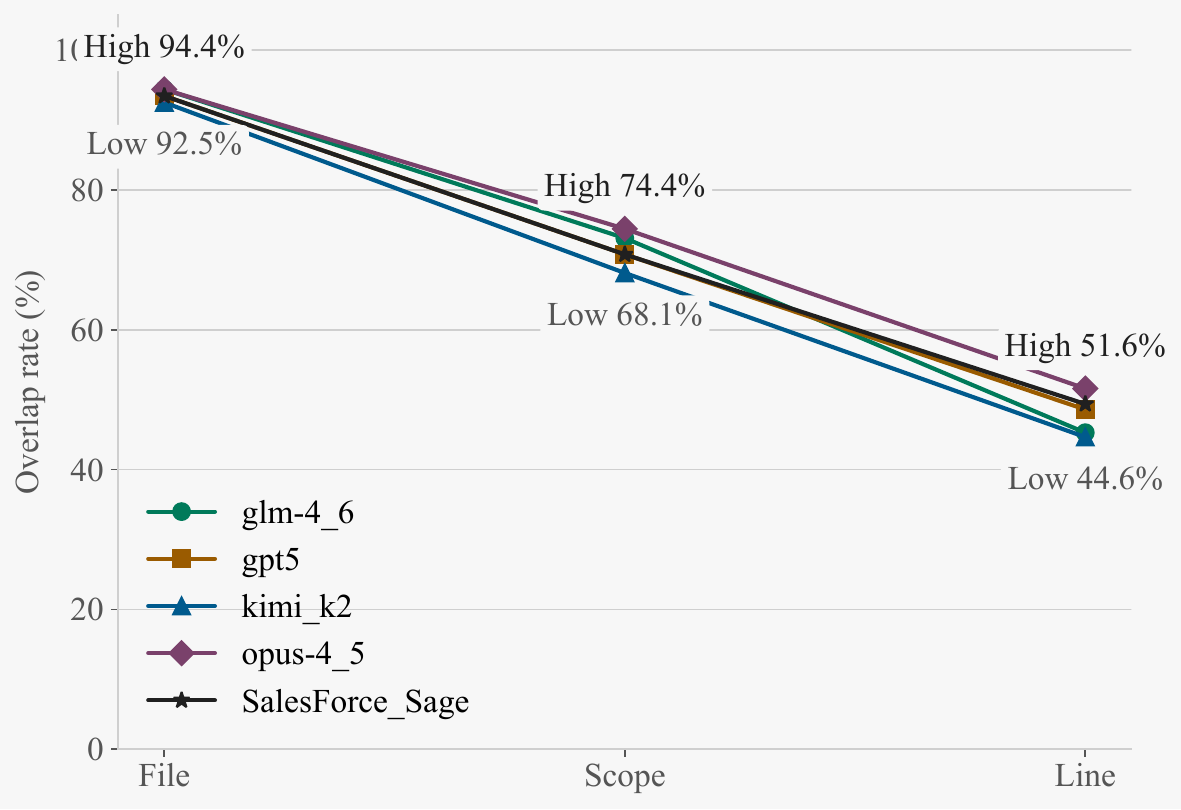}
  \caption{File-, scope-, and exact-line overlap across all required deletions
  in the 197 tasks solved by all five models}
  \label{fig:deletion-localization-elbow}
  \vspace{-6pt}
\end{figure}

\FloatBarrier

\subsection{Patch Size Comparison between Developer Patch and Model Generated Patch with Guard-and-Go}\label{patch-size}
\begin{table}[!htbp]
\centering
\small
\setlength{\tabcolsep}{4pt}
\renewcommand{\arraystretch}{1.05}
\caption{Model-generated patch size relative to the corresponding developer patch for passing Guard-and-Go pairs. LOC is measured as the sum of added and deleted lines.}
\label{tab:guard-and-go-size}

\begin{tabular}{lrrrr}
\toprule
\textbf{Model} &
\textbf{Pairs} &
\begin{tabular}[c]{@{}c@{}}\textbf{Model}\\\textbf{larger}\end{tabular} &
\begin{tabular}[c]{@{}c@{}}\textbf{Equal}\\\textbf{LOC}\end{tabular} &
\begin{tabular}[c]{@{}c@{}}\textbf{Model}\\\textbf{smaller}\end{tabular} \\
\midrule
GLM-4.6         &  91 &  89 (97.80\%) &  0 (0.00\%)  &  2 (2.20\%)  \\
GPT-5           & 120 &  58 (48.33\%) & 10 (8.33\%)  & 52 (43.33\%) \\
Kimi-K2         &  92 &  75 (81.52\%) &  3 (3.26\%)  & 14 (15.22\%) \\
Opus-4.5        &  91 &  30 (32.97\%) & 20 (21.98\%) & 41 (45.05\%) \\
Salesforce SAGE & 100 &  50 (50.00\%) & 14 (14.00\%) & 36 (36.00\%) \\
\midrule
\textbf{Overall} &
\textbf{494} &
\textbf{302 (61.13\%)} &
\textbf{47 (9.51\%)} &
\textbf{145 (29.35\%)} \\
\bottomrule
\end{tabular}
\end{table}
\FloatBarrier

% \begin{table}[!htbp]
% \centering
% \small
% \setlength{\tabcolsep}{3pt}
% \renewcommand{\arraystretch}{1.05}

% \begin{tabular}{lrrrrr}
% \toprule
% \textbf{Model} &
% \textbf{Tasks} &
% \begin{tabular}[c]{@{}r@{}}
% \textbf{Passed}\\
% \textbf{without}
% \end{tabular} &
% \begin{tabular}[c]{@{}r@{}}
% \textbf{Passed}\\
% \textbf{with}
% \end{tabular} &
% \begin{tabular}[c]{@{}r@{}}
% \textbf{Pass rate}\\
% \textbf{with}
% \end{tabular} &
% \textbf{Drop} \\
% \midrule
% GPT-5.6-Sol     & 34 & 21 & 15 & 44.12\% & 17.65 pp \\
% Opus-4.8        & 34 & 21 & 14 & 41.18\% & 20.59 pp \\
% GLM-5.2         & 34 & 26 & 18 & 52.94\% & 23.53 pp \\
% DeepSeek-V4-Pro & 34 & 18 & 10 & 29.41\% & 23.53 pp \\
% \midrule
% \textbf{Overall} &
% \textbf{136} &
% \textbf{86} &
% \textbf{57} &
% \textbf{41.91\%} &
% \textbf{21.32 pp} \\
% \bottomrule
% \end{tabular}

% \caption{Impact of adding the deletion-focused F2P test. ``Passed without''
% counts model-task attempts that would pass without the deletion-focused test.
% The final column reports the absolute reduction in pass rate in percentage
% points (pp).}
% \label{tab:deletion-f2p-impact}
% \end{table}
% \FloatBarrier

\section{Construction of the Guard-and-Go Taxonomy}
\label{app:grounded-theory}

We analyze every task--model pair that the strategy classifier in the main paper's ``Models Substitute Added Control Flow for Removal'' subsection labels Guard-and-Go.
In each pair, the model retained code that the official human patch deleted and added a guard that routes execution around the retained code.
The classifier had to quote both the retained lines and the added guard to justify its label, and we kept only pairs for which both quotes were recovered.
This process left 684 pairs for analysis.

\paragraph{Open and axial coding.}
Two models, MiniMax-M2.7 and Claude Sonnet~5, independently open-coded each pair after inspecting the problem statement, the official human-authored patch, the model-generated patch, and the classifier rationale.
Coding proceeded in six rounds of five pairs, covering 30 pairs in total.
After each round, an axial coder (Claude Opus 4.7) compared the new codes against the running taxonomy and added, merged, or revised categories.
Table~\ref{tab:axial-rounds} reports the result of each round.
The fifth and sixth rounds added no categories and left all 23 definitions unchanged, so we stopped coding and treated the taxonomy as saturated at 23 categories.

\begin{table}[t]
\centering
\small
\begin{tabular}{@{}lrrrr@{}}
\toprule
\textbf{Round} & \textbf{Pairs coded} & \textbf{Added} & \textbf{Revised} & \textbf{Total} \\
\midrule
1 &  5 & 12 & 0 & 12 \\
2 & 10 &  4 & 0 & 16 \\
3 & 15 &  4 & 0 & 20 \\
4 & 20 &  3 & 0 & 23 \\
5 & 25 &  0 & 0 & 23 \\
6 & 30 &  0 & 0 & 23 \\
\bottomrule
\end{tabular}
\caption{Categories in the running taxonomy after each axial round.
Each round covers five new pairs.}
\label{tab:axial-rounds}
\end{table}

\begin{table}[t]
\centering
\small
\setlength{\tabcolsep}{3pt}
\renewcommand{\arraystretch}{1.12}

\begin{tabularx}{\columnwidth}{
    >{\raggedright\arraybackslash}p{0.32\columnwidth}
    >{\raggedright\arraybackslash}X
}
\toprule
\textbf{Subtype} & \textbf{Definition} \\
\midrule
Retained Path as Live Fallback &
Guards the reported case but leaves logic deleted by the official patch as the default path, allowing non-matching inputs to continue executing the retained logic. \\

Special-Case Carve-Out Bypass &
Adds a branch or relaxes a precondition for one input category while leaving the general resolution mechanism unchanged. \\

Missing-Element Existence Bypass &
Checks for a required runtime element and returns a default value when it is absent instead of evaluating the retained logic. \\

Failure-Site Symptom Patch &
Suppresses or compensates for the symptom at the failure site while leaving the producing control flow unchanged. \\

Exception Capture Bypass &
Catches exceptions raised by retained logic and returns a default value instead of removing the underlying cause. \\

Post-Hoc Output Correction &
Corrects an incorrect result after the buggy computation has already completed. \\

Missing-Functionality Injection &
Introduces new logic or parameters to handle the reported case instead of adopting the restructuring performed by the official patch. \\

Upstream Reroute Around Retained Branch &
Redirects problematic inputs away from a retained branch, leaving the obsolete branch as unreachable legacy logic. \\

Parallel Duplicate-Path Retention &
Retains logic that the official patch consolidates while introducing an additional execution path that can also satisfy the behavior. \\

Value-Compatibility Shim &
Conditionally restores a legacy constant or decoding behavior for a specific input. \\
\bottomrule
\end{tabularx}

\caption{Guard-and-Go structural subtypes identified in model-generated patches.}
\label{tab:guard-and-go-subtypes}
\end{table}
\paragraph{Consolidation.}
Many of the 23 categories described the same behavior in different contexts.
For example, three categories described catching exceptions from the retained code and returning a default value.
They differed only in the number of exception types caught, so we consolidated them into one behavior, \emph{Exception Capture Bypass}, listed in Table~\ref{tab:gng-types}.
Five other categories described adding a branch for a special input and differed only in the input type: a property object, a regular expression, or a database join.
We consolidated them into one behavior, \emph{Special-Case Carve-Out Bypass}.
Applying this rule to all 23 categories yielded the ten types in Table~\ref{tab:gng-types}: six types merged two or more categories, while four remained unchanged.
Claude Opus~4.7 proposed the grouping, and the authors reviewed it and finalized the ten definitions before labeling any pairs.

We kept two similar types separate.
In \emph{Upstream Reroute Around Retained Branch}, the guard redirects the problematic input, so the retained code can no longer run.
In \emph{Retained Path as Live Fallback}, the retained code still runs for every input that the guard does not catch. Since the retained code is dead in the first case and live in the second, we report the types separately.

\begin{table}[!htbp]
\centering
\small
\setlength{\tabcolsep}{3.5pt}
\renewcommand{\arraystretch}{1.10}

\begin{tabular}{
    @{}
    >{\raggedright\arraybackslash}p{3.8cm}
    rrr
    @{}
}
\toprule
\textbf{Structural form} & \textbf{\(n\)} & \textbf{Share} & \textbf{Pass rate} \\
\midrule
Retained Path as Live Fallback
    & 221 & 40.2\% & 68.3\% \\

Special-Case Carve-Out Bypass
    &  95 & 17.3\% & 52.6\% \\

Missing-Element Existence Bypass
    &  60 & 10.9\% & 81.7\% \\

Failure-Site Symptom Patch
    &  51 &  9.3\% & 66.7\% \\

Exception Capture Bypass
    &  43 &  7.8\% & 53.5\% \\

Missing-Functionality Injection
    &  35 &  6.4\% & 74.3\% \\

Post-Hoc Output Correction
    &  20 &  3.6\% & 80.0\% \\

Upstream Reroute Around Retained Branch
    &  10 &  1.8\% & 90.0\% \\

Parallel Duplicate-Path Retention
    &   8 &  1.5\% & 87.5\% \\

Value-Compatibility Shim
    &   7 &  1.3\% & 71.4\% \\
\midrule
\textit{All typed pairs}
    & 550 & 100.0\% & 67.3\% \\
\bottomrule
\end{tabular}

\caption{Distribution of the ten Guard-and-Go structural forms. Of the 684
Guard-and-Go pairs, the closed-source classifier assigned a subtype to 550 and
abstained on the remaining 134. Share is computed over the 550 typed pairs, and
pass rate is the proportion of pairs that SWE-bench Verified records as
resolved.}
\label{tab:gng-types}
\end{table}

\paragraph{Closed coding.}
A final labeler (MiniMax-M2.7, temperature 0) applied the frozen taxonomy of ten Guard-and-Go types to all 684 pairs.
For each pair, the labeler selected the most specific type and cited the corresponding lines from the model-generated patch as evidence.
The labeler could also decline to assign a type. 134 task--model pairs were labeled ``other'', indicating behavior that the ten types do not name.
% Of these, 133 matched none of the ten definitions, which we interpret as cases where the classifier labeled the pair Guard-and-Go but the patch showed no guard around retained code.
Table~2 of the main paper and the accompanying analysis therefore cover the 550 pairs that received a type.

To check these labels, one author sampled five pairs from each of the ten types, for 50 pairs in total, and confirmed that the cited lines exhibited the behavior described by the assigned definition.
This validation assesses whether the labels match the taxonomy's structural definitions; it does not evaluate each patch's correctness, maintainability, or intent. We provide the definition of each subtype in~\ref{tab:guard-and-go-subtypes}.

\section{Construction of Deletion-Focused F2P Tests}
\label{app:deletion-f2p}
We retrofit selected SWE-bench Verified tasks with deletion-focused \texttt{FAIL\_TO\_PASS} (F2P) tests. The construction process first identifies a behaviorally important deletion that is not covered by the existing visible tests and then creates a repository-native test that fails when the deleted code remains and passes when the deletion is performed.

\subsection{Selecting Behaviorally Important Deletions}
\label{app:f2p-target-selection}

We begin with 69 tasks for which deletions constitute at least 25\% of the changed lines in the developer patch. Our analysis covers 65 of these tasks. Because deleted lines may include comments, blank lines, docstrings, delimiters, and other structurally trivial changes, we do not treat every deleted line as an independent target. Instead, we analyze the source version preceding the developer patch and group related deleted lines into deletion units. For Python files, the source structure is used to associate deleted lines with their enclosing syntax elements. When the source cannot be parsed, nearby deleted lines are grouped into a single unit.

A deletion unit may represent a removed condition, control-flow statement, function signature, assignment, call, decorator, or complete compound statement such as an \texttt{if}, \texttt{for}, \texttt{try}, function, or class block. Imports are considered meaningful only when they are removed together with the code that depends on them. This process identifies 166 deletion units across 63 tasks; the remaining two tasks contain no meaningful code deletion after non-code and structurally trivial lines are excluded.

We rank the deletion units using the heuristic in Table~\ref{tab:deletion-importance-scoring}. The score prioritizes deletions that alter control flow, remove complete program constructs, correspond to entities mentioned in the problem statement, and disappear entirely from the developer patch. Standalone import removals receive a negative score because they generally reflect cleanup induced by another deletion rather than the primary behavior required by the task. The highest-scoring unit is selected as the deletion target for each task.

\begin{table}[t]
\centering
\small
\setlength{\tabcolsep}{3pt}
\renewcommand{\arraystretch}{1.05}
\caption{Heuristic used to prioritize behaviorally important deletion units.}
\label{tab:deletion-importance-scoring}
\begin{tabular}{p{0.76\columnwidth}r}
\toprule
\textbf{Signal} & \textbf{Score} \\
\midrule
Deleted lines are not reintroduced elsewhere &  $+3.0$ \\
Complete control block is removed & $+3.0$ \\
Condition guarding existing code is changed or removed & $+3.0$ \\
Control-flow statement is removed & $+3.0$ \\
Complete function or class is removed & $+2.5$ \\
Decorator is removed & $+2.0$ \\
Comparison or Boolean logic is removed & $+1.5$ \\
Function or class signature is changed & $+1.5$ \\
Assignment is removed & $+1.0$ \\
Function or method call is removed & $+1.0$ \\
Deleted code contains problem-statement terms &  $+2.0$ \\
Larger multi-line deletion unit &  $+1.5$ \\
Import and its dependent code are removed & $+0.5$ \\
Standalone import is removed & $-4.0$ \\
\bottomrule
\end{tabular}
\end{table}

For patch-level analysis, we consider a model to have adopted the selected deletion when it removes at least half of the lines belonging to that unit. Among the 63 selected units, 24 are covered by a listed visible test. The remaining 39 form the candidate pool for deletion-focused F2P construction.

\begin{table}[t]
\centering
\small
\setlength{\tabcolsep}{4pt}
\renewcommand{\arraystretch}{1.05}
\caption{Task counts across the deletion-focused F2P construction process.}
\label{tab:deletion-f2p-selection}
\begin{tabular}{p{0.78\columnwidth}r}
\toprule
\textbf{Selection stage} & \textbf{Count} \\
\midrule
Tasks with at least 25\% deletion & 69 \\
Tasks included in deletion-unit analysis & 65 \\
Tasks with a meaningful code deletion & 63 \\
Tasks without a meaningful code deletion & 2 \\
Deletion units identified & 166 \\
Targets covered by a listed visible test & 24 \\
Targets not covered by a listed visible test & 39 \\
Host-validated deletion-focused F2P tasks & 34 \\
\bottomrule
\end{tabular}
\end{table}

\subsection{Retrofitting and Validating F2P Tests}
\label{app:f2p-test-construction}

For each of the 39 candidates, we first establish a source-level oracle that distinguishes the base version from the developer-patched version. The selected deletion must be present in the base source and absent from the corresponding location after applying the developer patch. Consequently, the resulting predicate fails on the base version because the targeted code remains and passes on the gold version because the intended deletion has occurred.

The test reads the target source file from the current checkout and compares normalized source lines. It does not import or execute the target project code. This design isolates the structural edit from project initialization, database fixtures, framework settings, and other runtime side effects. When the selected lines disappear entirely from the developer-patched file, the predicate checks their absence from the complete file. When the same text is relocated, rewritten, or duplicated elsewhere, the check is restricted to a stable enclosing function or class. A candidate is excluded if no stable scope can distinguish the base and gold versions.

Each test is placed within the repository's existing test organization. For Django projects, the test is added as a \texttt{unittest} module under the same test application used by the original task. For SymPy, it is expressed as a plain test function compatible with the repository's selector format. For other pytest-based repositories, the test file is placed alongside the original task's test location. The derived SWE-bench record contains only the deletion-focused test in its \texttt{test\_patch}, lists the generated selector under \texttt{FAIL\_TO\_PASS}, and leaves \texttt{PASS\_TO\_PASS} empty. This isolates the evaluation outcome from the task's original visible tests.

We validate every generated test against materialized base and gold source trees. Retention requires the test to be collected correctly, fail on the base version with the intended deletion assertion, and pass on the gold version. Tests that error, skip, are not collected, or fail to distinguish the two versions are discarded. We refer to tests satisfying these conditions as \emph{host validated}; \emph{harness validated} additionally denotes execution within the complete SWE-bench evaluation environment. Of the 39 candidates, 34 satisfy host validation. The remaining five are excluded because no stable source-level predicate separates the base and gold versions.

\section{Extended \textsc{CanItDelete} Design and Results}
\label{app:canitdelete-design}

\subsection{Candidate Mining and Structural Ranking}

\paragraph{Repository population.}
For each of Python and JavaScript, we select the 100 most-starred active,
non-fork public repositories and traverse their reachable non-merge histories.
Activity after January 1, 2024 defines the repository population; older
commits within those repositories remain eligible.

\paragraph{Deletion-only file edits.}
For each commit, we reconstruct every changed file before and after the edit.
We retain Python, JavaScript, JSX, and JavaScript-module edits that add no
lines, have a pre-edit file of at most 100~KB, have recoverable and decodable
revisions, and leave a nonempty post-edit file. We exclude binary changes and
file deletions but retain test files. Selection is file-level, so other files
in the commit may contain additions or modifications. This yields 79,074
eligible edits, each with its Git-derived post-edit file as the reference.

\paragraph{Structural challenge index.}
Because \textsc{CanItDelete} is a diagnostic stress test rather than a
representative GitHub sample, we rank candidates by three established
change-complexity dimensions: pre-edit code size, change size, and dispersion
\cite{nagappan2005relative,hassan2009predicting,kamei2013large,moser2008comparative}.
We use the dimensions, but not defect-prediction coefficients, because defect
risk and editing difficulty are different constructs.

Let $L$ be the number of nonempty pre-edit lines, $C$ the number of nonempty
source lines deleted by the reference edit, and $H$ the number of deletion
hunks in a unified diff with three lines of context. Removing duplicate
repository--commit--file identities leaves $N=74{,}485$ candidates. For
$k\in\{L,C,H\}$, we compute the right-continuous empirical percentile
\begin{equation}
p_k(x)=\frac{1}{N}\sum_{i=1}^{N}\mathbf{1}[x_{ik}\leq x_k]
\label{eq:app-canitdelete-percentile}
\end{equation}
and the equal-weight structural challenge index
\begin{equation}
d_{\mathrm{V2}}=\frac{p_L+p_C+p_H}{3}.
\label{eq:app-canitdelete-score}
\end{equation}
Percentiles avoid distributional assumptions and hand-selected saturation
thresholds. Equal weights avoid imposing an unsupported exchange rate among
file reading, deletion volume, and multi-site coordination. Thus,
$d_{\mathrm{V2}}$ ranks structural challenge; it is not a calibrated failure
probability.

\paragraph{Final task selection.}
We retain the highest-scoring file per repository--commit pair, preventing one
logical change from occupying multiple positions, and select the top 200
edits without using model performance. The benchmark spans 200 commits from
35 repositories: 151 Python and 49 JavaScript-family tasks, including 53 test
files. Every task has at least three separated deletion hunks.

\begin{figure}[t]
    \centering
    \includegraphics[width=\columnwidth]
    {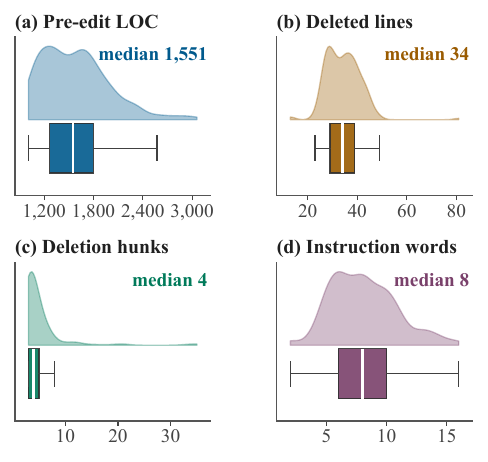}
    \caption{Structural distributions across the 200 \textsc{CanItDelete}
    tasks.}
    \label{fig:app-canitdelete-profile}
\end{figure}

Figure~\ref{fig:app-canitdelete-profile} summarizes the benchmark. Median
(IQR) values are 1,551 (1,266--1,800) pre-edit LOC, 34 (29--39) deleted lines,
4 (3--5) deletion hunks, and 8 (6--10) instruction words.

\subsection{Instruction Generation and Task Validation}\label{app:instruction-gen-task-val}

After structural selection, GPT-5.6 Sol receives the pre-edit file, reference
diff, file path, and commit message as weak context. It generates a short,
present-tense request covering the complete deletion without additions,
invented motivation, or references to a patch or diff. Mechanical checks
reject malformed requests, non-deletion scope, prohibited language, and
patch-oriented terms. The replication package provides the full prompts.

Each task passes an LLM gate and an author gate. Given only the instruction and
pre-edit file, the LLM solver view checks nontriviality, grounding, and whether
all edit sites are locatable. Given the full before/after record, its evaluator
view checks faithfulness, multi-site consistency, reference correctness,
scope, and cross-field coherence. The author then verifies that the request
covers every substantive Git deletion without broadening the task and that
all sites and boundaries are recoverable without the hidden diff, post-edit
file, repository context, or tests. Inventory-like requests are shortened and
revalidated. This separation of Git provenance, generated descriptions, and
deterministic evaluation follows recent repository-derived benchmarks
\cite{gittaskbench,dscodebench,domaineval,latesteval}.

\subsection{Diagnostic Modes}
\label{app:canitdelete-modes}

The four modes are cumulative: each retains prior guidance and adds one
controlled signal. Changes between stages identify missing information;
failure with exact lines isolates execution and scope control after intent,
region, and boundaries are supplied.

\begin{table*}[t]
    \centering
    \small
    \caption{The four cumulative \textsc{CanItDelete} diagnostic modes.}
    \label{tab:app-canitdelete-modes}
    \begin{tabular}{p{0.16\textwidth}p{0.40\textwidth}p{0.36\textwidth}}
        \toprule
        Mode & Cumulative signal & Diagnostic question \\
        \midrule
        Vanilla
            & No signal beyond the developer-style request
            & Can the model infer and execute the complete deletion end to
              end? \\
        Explicit deletion
            & Requires complete removal and forbids guards, comments,
              disabled branches, fallbacks, replacement code, and other
              workarounds
            & Is failure caused by not choosing a genuinely subtractive
              edit? \\
        Region pointer
            & Identifies relevant functions, methods, or regions without
              revealing deletion boundaries
            & Is finding every required region the principal obstacle? \\
        Exact lines
            & Supplies occurrence-specific spans and requires all unmarked
              text to be preserved
            & Can the model execute a precise deletion once intent, location,
              and boundaries are known? \\
        \bottomrule
    \end{tabular}
\end{table*}

\subsection{Deletion-Compliance Evaluator}
\label{app:canitdelete-evaluator}

No LLM judges evaluate outputs. An output is deletion-compliant when the full
target is absent, the remaining executable structure agrees with the
reference, and no behavior-affecting directive or unrelated code change is
introduced. Formatting, whitespace, and ordinary explanatory comments are
accepted; commenting out or disabling the target is not. Type-checking, lint,
and coverage directives count when they can affect behavior. Exact reference
matches are always accepted, including for unsupported source dialects.

The evaluator derives occurrence-specific units from the pre-edit-to-reference
diff. Source coordinates, hunk membership, and anchors prevent one occurrence
of a repeated line from receiving credit for another. Decisions combine
hunk-anchored alignment, Python AST or JavaScript parser structure, token and
dialect fallbacks, and raw-source checks for targets preserved in comments,
literal-false branches, early-return wrappers, or similar nonexecuting forms.

Failures are partitioned by whether required code remains. \emph{Incomplete
deletion} covers partial removal, additions while target code remains, mixed
missing and excess deletion, wrong-site or wrong-boundary deletion, no change,
disabled or commented targets, and syntax-invalid outputs retaining the
target. \emph{Complete deletion, invalid edit} covers over-deletion,
scope-changing edits, and syntax-invalid outputs after target removal. The
partition separates incomplete removal from failure to preserve its boundary.

\subsection{Complete Diagnostic-Ladder Results}
\label{app:canitdelete-ladder-details}

GLM-5.2, GPT-5.6 Sol, MiniMax-M3, and Qwen3-235B use all 200 tasks
in every mode. Provider failures left Claude Opus 4.8 with 173 tasks having
usable responses in all four modes, so its trajectory uses that paired subset.
Each within-model comparison therefore holds its task set fixed; only the
Opus panel has fewer tasks than the vanilla analysis.
Table~\ref{tab:app-canitdelete-ladder} reports the complete outcome partition.

\begin{table*}[t]
    \centering
    \scriptsize
    \setlength{\tabcolsep}{5pt}
    \caption{Complete five-model diagnostic-ladder results. Four models use
    200 tasks per mode; Claude Opus 4.8 uses 173. Cells report count
    (percentage).}
    \label{tab:app-canitdelete-ladder}
    \begin{tabular}{lrrr}
        \toprule
        Mode & Success & Incomplete deletion
             & Complete deletion, invalid edit \\
        \midrule
        \multicolumn{4}{@{}l}{\textit{GLM-5.2}} \\
        Vanilla          & 133 (66.5) & 30 (15.0) & 37 (18.5) \\
        Explicit deletion& 128 (64.0) & 36 (18.0) & 36 (18.0) \\
        Region pointer   & 147 (73.5) & 29 (14.5) & 24 (12.0) \\
        Exact lines      & 175 (87.5) &  6 (3.0)  & 19 (9.5) \\
        \midrule
        \multicolumn{4}{@{}l}{\textit{GPT-5.6 Sol}} \\
        Vanilla          & 148 (74.0) & 20 (10.0) & 32 (16.0) \\
        Explicit deletion& 144 (72.0) & 21 (10.5) & 35 (17.5) \\
        Region pointer   & 148 (74.0) & 19 (9.5)  & 33 (16.5) \\
        Exact lines      & 161 (80.5) &  6 (3.0)  & 33 (16.5) \\
        \midrule
        \multicolumn{4}{@{}l}{\textit{MiniMax-M3}} \\
        Vanilla          & 134 (67.0) & 36 (18.0) & 30 (15.0) \\
        Explicit deletion& 134 (67.0) & 32 (16.0) & 34 (17.0) \\
        Region pointer   & 137 (68.5) & 33 (16.5) & 30 (15.0) \\
        Exact lines      & 170 (85.0) &  6 (3.0)  & 24 (12.0) \\
        \midrule
        \multicolumn{4}{@{}l}{\textit{Claude Opus 4.8}} \\
        Vanilla          & 136 (78.6) & 25 (14.5) & 12 (6.9) \\
        Explicit deletion& 138 (79.8) & 27 (15.6) &  8 (4.6) \\
        Region pointer   & 143 (82.7) & 19 (11.0) & 11 (6.4) \\
        Exact lines      & 169 (97.7) &  1 (0.6)  &  3 (1.7) \\
        \midrule
        \multicolumn{4}{@{}l}{\textit{Qwen3-235B-A22B}} \\
        Vanilla          &  50 (25.0) & 109 (54.5) & 41 (20.5) \\
        Explicit deletion&  55 (27.5) & 105 (52.5) & 40 (20.0) \\
        Region pointer   &  54 (27.0) &  99 (49.5) & 47 (23.5) \\
        Exact lines      & 113 (56.5) &  35 (17.5) & 52 (26.0) \\
        \bottomrule
    \end{tabular}
\end{table*}

Explicit deletion changes success by only $-2.5$ to $+2.5$ percentage points
relative to vanilla. Region pointers change success by 0.0--7.0 points, with
the largest gain for GLM-5.2. Only exact lines improve all five models, by
6.5--31.5 points. Incomplete deletion then falls to 0.6--3.0\% for four
models, while Qwen3-235B retains target code in 17.5\% of tasks.

The remaining failures expose a separate boundary-control deficit. Claude
Opus 4.8 ends with 1.7\% complete-deletion invalid edits, but GPT-5.6 Sol
remains at 16.5\%. For Qwen3-235B, this rate rises from 20.5\% to 26.0\% as
incomplete deletion falls. Exact localization can therefore replace omitted
deletions with edits that remove the target but change too much. The ladder
separates finding every target occurrence from stopping at its boundary.

\section{Deletion-Focused Training Data Selection}
\label{app:sft-selection}

\subsection{File-level examples.}
Unlike \textsc{CanItDelete}, which retains the most structurally difficult edits, we sample across repositories, languages, and deletion difficulty. Given an instruction and the complete pre-edit file, DeepSeek-V3.2 generates
the complete edited file. We apply deterministic rejection sampling using the criteria in Section~\ref{subsec:benchmark_design}, yielding 10,000 accepted responses whose required deletions are confined to one file.

\subsection{Repository-level examples.}
We retain non-root, non-merge commits whose Python changes contain only deletions, touch at least two files, and span 3--1,500 changed lines. For each commit, we generate an instruction and an F2P test that fails before the reference edit and passes afterward, and package the task in Harbor format \cite{merrill2026terminal}. We use mini-SWE-agent with MiniMax-M2.7 as the teacher and apply rejection sampling based on generated F2P tests, obtaining 2,821 samples for training.
  
\end{document}